# Photonic Chern insulator through homogenization of an array of particles

Meng Xiao and Shanhui Fan[*]

[1]Department of Electrical Engineering, and Ginzton Laboratory, Stanford University, Stanford, California 94305, USA

[*]Corresponding E-mail: shanhui@stanford.edu

Abstract: We propose a route towards creating a metamaterial that behaves as a photonic Chern insulator, through homogenization of an array of gyromagnetic cylinders. We show that such an array can exhibit non-trivial topological effects, including topologically non-trivial band gaps and one-way edge states, when it can be homogenized to an effective medium model that has the Berry curvature strongly peaked at the wavevector $k=0$. The non-trivial band topology depends only on the parameters of the cylinders and the cylinders' density, and can be realized in a wide variety of different lattices including periodic, quasi-periodic and random lattices. Our system provides a platform to explore the interplay between disorder and topology and also opens a route towards synthesis of topological meta-materials based on the self-assembly approach.

The classification of band structure in terms of band topology in classical waves has led to the newly emerging field of topological photonics [1,2] and phononics [3]. Numerous realizations of topological nontrivial systems in different classical waves have been proposed and experimentally verified [4-32] after the initial theoretical prediction of one-way edge mode in a photonic crystal with a nontrivial band topology [4]. Recent progress also includes the investigation of classical analogue of three-dimensional (3D) topological insulators [24,25] and semimetals [26,27]. The vast majority of these works use periodic lattices, either explicitly as in topological photonic crystal and phononic crystals, or implicitly in topological meta-materials, where in order to achieve a prescribed effective permittivity and permeability, the underlying physical structure is typically periodic.

Complementing the works on periodic systems, in this Letter, we propose a route towards creating a meta-material that behaves as a photonic Chern insulator through homogenization of an array of gyromagnetic particles as shown in Fig. 1(a). The construction of such a physical metamaterial structure is motivated by its corresponding effective medium model that supports a non-trivial topological band gap. For any physical metamaterial system consisting of a periodic array of sub-wavelength elements, its effective medium model is only valid near the center of the Brillouin zone. Yet topological properties are global properties across the entire Brillouin zone. Therefore, even though non-trivial topological effects have been noted in uniform systems with certain bulk parameters, [8,28-31], in general it is not obvious, *a priori*, that the topology of a physical metamaterial system consisting of an array of sub-wavelength elements can be understood in terms of its effective medium model, especially when such an effective medium model is local [28]. In our case, the key idea is to construct an effective medium in which the Berry curvature of its band structure is strongly peaked at $k=0$, where $k$ is the wavevector. Since the effective medium model should describe the physical structure very well near $k=0$, we expect that the non-trivial band topology of the effective medium model should also manifest in the physical metamaterial systems. And indeed, through numerical simulations, we find that our metamaterial systems with a wide variety of lattices, including periodic, quasiperiodic, and random lattices, all of which homogenize to the same effective medium model, all possesses complete non-trivial topological band gap.

From a fundamental physics perspective, the systems that we consider here may provide a platform to explore the interplay between order or disorder and topology. From an experimental perspective, creating a disordered system with non-trivial topology may relax the stringent requirements for fabricating topologically non-trivial photonic and phononic systems. While there have been several very recent works on nontrivial topological photonic structures utilizing aperiodic lattices [33-35], our construction differs in that it is not based on any specific lattice [33,34] or needs to engineer the local connections [35]. The structure reported here may therefore open a route towards synthesis of topological meta-material based on the self-assembly approach.[36-38]

We first introduce an effective medium model that exhibits non-trivial topology in its band structure, and has the Berry curvature of its band structure strongly peaked at $k=0$. We consider electromagnetic waves in two dimensions and focus on the TM polarization, which has the electric field $E_z$ along the $z$-direction and the magnetic field $H_x$, $H_y$ in the $xy-$ plane. For such a TM polarized wave, we consider an effective medium model where the relevant effective electromagnetic materials are the relative electric permittivity $\varepsilon_e$ along the $z$-direction, and the relative magnetic permeability in the $xy$-plane, which has the form

$$\vec{\mu}_e = \begin{pmatrix} \mu_r & i\mu_k \\ -i\mu_k & \mu_r \end{pmatrix}. \tag{1}$$

For this system, the Maxwell equations for a plane wave at a frequency $\omega$ can be written as

$$\begin{pmatrix} \omega\mu_r\mu_0 & i\omega\mu_k\mu_0 & -k_y \\ -i\omega\mu_k\mu_0 & \omega\mu_r\mu_0 & k_x \\ -k_y & k_x & \omega\varepsilon_e\varepsilon_0 \end{pmatrix} \begin{pmatrix} H_x \\ H_y \\ E_z \end{pmatrix} = 0, \tag{2}$$

where $\varepsilon_0$ and $\mu_0$ are the permittivity and permeability of vacuum, respectively, $k_x$ and $k_y$ are the wavevectors, respectively.

In order to achieve a band structure with non-trivial topology, we start with the case where $\mu_k = 0$, and

$$\begin{aligned} \varepsilon_e &= \gamma_\varepsilon (\omega - \omega_E)/\omega_E \\ \mu_r &= \gamma_\mu (\omega - \omega_H)/\omega_H \end{aligned}. \tag{3}$$

The starting point of our system is therefore the Epsilon and Mu Near Zero (EMNZ) systems that have been widely considered in the meta-material literature. [39] Here we consider a lossless system. $\gamma_e$ and $\gamma_\mu$ are both positive as required by causality. [40] When $\omega_E \neq \omega_H$, (Fig. 2(a)), the system supports a mode that is singly degenerate at $k=0$ having a frequency $\omega = \omega_E$. This mode exhibits a quadratic dispersion at $k \neq 0$. The system also supports at $k=0$ a pair of doubly degenerate modes at $\omega = \omega_H$. At $k \neq 0$ the two modes split into two bands, one with flat dispersion and the other with quadratic dispersion. By setting $\omega_E = \omega_H \equiv \omega_0$, we force a three-fold accidental degeneracy at $k=0$. [41,42] At $k \neq 0$, the three modes split into two bands with a Dirac-like linear dispersion as well as a flat band, as shown in Fig. 2(b).

Starting with the band structure in Fig. 2(b), we then break time-reversal symmetry by setting a frequency-independent $\mu_k \neq 0$. Now the three-fold degeneracy at $k=0$ is completely lifted as shown in Fig. 2(c) and two band gaps are introduced into the system. Both of these band gaps are topologically non-trivial. To see the nontrivial topology, here we develop a Hamiltonian for the Maxwell equations near $\omega_0$. For the system with $\mu_k = 0$, an effective Hamiltonian has been developed to describe the physics in the vicinity of the triply degenerate point. [44]. For our system here with $\mu_k \neq 0$, starting from Eq. (2), and keeping only the lowest order of $\Delta\omega \equiv \omega - \omega_0$ we obtain

$$\begin{pmatrix} 0 & -i\tilde{\mu}_k & \tilde{k}_y \\ i\tilde{\mu}_k & 0 & -\tilde{k}_x \\ \tilde{k}_y & -\tilde{k}_x & 0 \end{pmatrix} \begin{pmatrix} \tilde{H}_x \\ \tilde{H}_y \\ \tilde{E}_z \end{pmatrix} = \frac{\Delta\omega}{\omega_E} \begin{pmatrix} \tilde{H}_x \\ \tilde{H}_y \\ \tilde{E}_z \end{pmatrix}, \qquad (4)$$

where $\tilde{\mu}_k = \mu_k / \gamma_\mu$, $\tilde{k}_{x,y} = k_{x,y} c / (\omega_0 \sqrt{\gamma_\varepsilon \gamma_\mu})$, $\tilde{H}_{x,y} = \sqrt{\mu_0 \gamma_\mu} H_{x,y}$, $\tilde{E}_z = \sqrt{\varepsilon_0 \gamma_\varepsilon} E_z$ and $c$ is the speed of light. The matrix on the right-hand side of Eq. (4) can now serve as an effective Hamiltonian. The three eigenvalues of this Hamiltonian are given by $\left\{0, \pm\sqrt{\tilde{k}_x^2 + \tilde{k}_y^2 + \tilde{\mu}_k^2}\right\}$, which is consistent with the band dispersion in Fig. 2(c). With the effective Hamiltonian obtained, we can determine the total Berry flux for the upper, middle, and lower bands to be $-\text{sgn}(\mu_k) 2\pi$, $0$ and $\text{sgn}(\mu_k) 2\pi$, respectively, when $\mu_k \neq 0$. Thus we have introduced an effective medium model with topologically non-trivial band gaps. Such an effective medium model thus represents a photonic Chern insulator. While there have been many theoretical proposals and experimental demonstrations of photonic Chern insulators [4-9], the effective medium model as proposed here represents a route for creating a Chern insulator that was not previously reported. In this model, the Berry curvature peaks at $k=0$, as shown in Fig. 2(d). Therefore, we anticipate that this effective medium model can be used to guide the construction of physical meta-material structures through homogenization.

Motivated by the effective medium model as presented above, we now consider physical metamaterial systems which homogenize to this model at $k=0$. We first consider an individual cylinder with radius $r_c$, relative permittivity $\varepsilon_c$ and the relative permeability in the $xy$-plane

$$\vec{\vec{\mu}}_c = \begin{pmatrix} 1 & i\kappa \\ -i\kappa & 1 \end{pmatrix}, \tag{5}$$

where $\kappa$ is assumed to be frequency independent. This is a simplified model for gyromagnetic effects. A more sophisticated model of the gyromagnetic effects can be found in Ref.[43]. The main results of the paper are not affected by the use of the more sophisticated model. The electric and magnetic dipole responses of a cylinder are described by

$$\begin{cases} \vec{p} = \varepsilon_0 \alpha_E \vec{E}^{\text{loc}} \\ \vec{m}_\pm = \alpha_\pm \vec{H}^{\text{loc}} \end{cases}, \tag{6}$$

where $\vec{E}^{\text{loc}}$ and $\vec{H}^{\text{loc}}$ represent the local electric and magnetic fields, respectively, $\alpha_E = 4iD_0/k_0^2$ and $\alpha_\pm = 8iD_{\pm 1}/k_0^2$ represent respectively, the electric and magnetic dipole polarizabilities, $k_0$ is the wavevector in vacuum and

$$D_n = \frac{(1-\kappa^2)k_0 r_c J_n'(k_0 r_c) J_n(k_c r_c) - J_n(k_0 r_c)\left[n\kappa J_n(k_c r_c) + k_c r_c J_n'(k_c r_c)\right]}{(1-\kappa^2)k_0 r_c H_n^{(1)\prime}(k_0 r_c) J_n(k_c r_c) - H_n^{(1)}(k_0 r_c)\left[n\kappa J_n(k_c r_c) + k_c r_c J_n'(k_c r_c)\right]}. \tag{7}$$

Here $k_c = \sqrt{\varepsilon_c \sqrt{1-\kappa^2}} k_0$, $J_n(x)$, $H_n^{(1)}(x)$, $J_n'(x)$ and $H_n^{(1)\prime}(x)$ are the Bessel function and Hankel function of the first kind and their derivatives, respectively.

We then consider a metamaterial system consisting of an array of cylinders as described above. An example of such metamaterial is shown in Fig. 1(a). Following the procedure in Ref. [44], we derive the effective medium parameters by solving the scattering problem as illustrated in Fig. 1(b). Here one particle is set at the center of a cylindrical cavity filled with air and surrounded by a background consisting of the effective medium. The radius of the cylindrical cavity $r_0$ is chosen such that $\pi r_0^2 / a^2 = 1$, where $1/a^2$ is the number of cylinders inside a unit area for the random array. The parameters of the effective medium are determined by assuming that there is no scattering by the cavity for waves incident from the effective medium in the limit where the wavelength in the effective medium is much larger than $r_0$, in which case only the electric and magnetic dipole responses of the cylinder need to be taken into account. The no-scattering condition gives (Refer to the Supplemental Material S-I for detailed derivations.)

$$\varepsilon_e = -\frac{2\left[4J_0'(k_0 r_0) + i\alpha_E k_0^2 H_0^{(1)\prime}(k_0 r_0)\right]}{k_0 r_0 \left[4J_0(k_0 r_0) + i\alpha_E k_0^2 H_0^{(1)}(k_0 r_0)\right]}, \quad (8)$$

and

$$\mu_r' \mp \mu_k' = \frac{k_0 r_0 \left[8J_1'(k_0 r_0) + i\alpha_\pm k_0^2 H_1^{(1)\prime}(k_0 r_0)\right]}{\left[8J_1(k_0 r_0) + i\alpha_\pm k_0^2 H_1^{(1)}(k_0 r_0)\right]}, \quad (9)$$

with $\mu_k = -\mu_k' / (\mu_r'^2 - \mu_k'^2)$, $\mu_r = \mu_r' / (\mu_r'^2 - \mu_k'^2)$. It can be proved that the effective parameters as described in Eqs. (8) and (9) are purely real when the system is nonabsorptive. [44] (See also Supplemental Material S.I) Note here the effective electric permittivity $\varepsilon_e$ and magnetic permeability $\mu_e$ are controlled by the electric and magnetic dipole responses of the cylinders independently. When the time-reversal symmetry is preserved, i.e, when $\kappa = 0$ in Eq. (5), the effective parameters having the form of Eq. (3) can be achieved by choosing the parameters of the cylinders such that the electric and magnet dipole responses have the same resonant frequencies, which then results in the dispersion relation shown in Fig. 2(b). [41] Starting from such a cylinder, the dispersion relation in Fig. 2(c) can then be achieved by setting $\kappa \neq 0$ in Eq. (5), which breaks the time-reversal symmetry.

Using full wave simulations, we now show that different lattices of the cylinders as discussed above,

all of which homogenize to the same effective medium model has described above, in fact all possesses non-trivial topology in their band structures. In Fig. 3(a), we consider the square lattice case. The cylinders have the parameters $r_c = 0.1735a$, $\varepsilon_c = 20$, and $\kappa = 0$, which are determined following the homogenization procedure as outlined above. The three-fold degeneracy at the $\Gamma$ point (i.e. $k = 0$ point) and conical-like dispersion near the $\Gamma$ point are consistent with the effective model as plotted in Fig. 2(b). In fact, we can find similar band dispersions near the $\Gamma$ point as those in Fig. 2(a)-(c) by varying some of the parameters of the cylinders. (See the Supplemental Material S. II). Based on the square lattice discussed above, we now consider the strip geometry shown in Fig. 3 (c). The lattice consists of the same cylinder as discussed above but with $\kappa = 0.08$. The lattice is periodic along the *y* direction and truncated by a perfect magnetic conductor (PMC) boundary on the left and a perfect electric conductor (PEC) boundary on the right. The projected band and the corresponding surface states are shown in Fig. 3(b). The field distribution of one of the surface states near the PEC boundary is also shown in Fig. 3(c). The projected band structure consists of three groups of bulk bands. Dispersions of the upper and lower groups of bulk bands agree well with the effective medium model near the $\Gamma$ point. And as predicted by the effective medium model, the breaking of time reversal symmetry introduces a local Berry flux of $2\pi$ and one-way surface states emerge near the $\Gamma$ point.

On the other hand, the effective medium model does not describe the band structure for the wavevectors significantly away from the $\Gamma$ point. First, the middle band is no longer perfectly flat. Second, as shown in Fig. 3a, there exists a band degeneracy between the lower two bands at the M point protected by the combination of time-reversal symmetry and the $C_{4v}$ lattice symmetry. Such degeneracy is not predicted by our effective medium model. When this degeneracy is lifted by breaking the time-reversal symmetry, a Berry flux of $2\pi$ peaks at the M point and as a result, the lower band becomes topologically trivial. The surface states inside the lower band gap, emerge near the $\Gamma$ point, almost reach the other band, but then merge back into their original bands.

In the square lattice structure, we observe that the effective medium model of Fig. 2 generally agrees well with the band structure of the physical system near $\Gamma$ point. Moreover, the upper band of Fig. 2 in the effective medium model agrees quite well with the band structure of the physical system. As a result one can achieve a topologically non-trivial band gap between the upper and the middle band in the physical systems. And a one-way edge state can be found in such a band gap in a truncated lattice. The lower band, on the other, significantly differ from the effective medium model away from $\Gamma$. These observations turn out to be generally applicable for other lattices that homogenize to the same effective medium model. As an additional example, results for a triangular lattice are shown in Supplemental Materials S. III. Therefore, we have shown that the effective medium model as we developed here provides the guidance for creating a class of non-trivial topological meta-material with different periodic lattices.

We now show that the effective medium model can be used to guide us in constructing random or quasi-periodic system with a non-trivial topological band gap and a one-way edge state. As an exemplary demonstration of random systems, we consider a geometry with a supercell as shown in Fig. 1(a). The supercell forms a square lattice and each supercell contains 25 cylinders with the same radius, dielectric constant and density of cylinders as in Fig. 3(b), except with $\kappa = 0.4$. Within each supercell, the cylinders are randomly distributed with equal probability while keeping the minimal distances between cylinders to be larger than $0.8a$ so that the dipole approximation in Eq. (6) remains valid. The projected band structure for a strip geometry with 5 supercells along the *y* direction is shown in Fig. 4a. The strip is truncated in the *y*-direction with PEC boundaries on the upper and lower sides, and is assumed to be periodic along the *x* direction. The structure supports a topological non-trivial band gap, the frequency range of which corresponds to the upper band gap in the effective medium model. Within the gap the structure supports one-way surfaces states with opposite propagation direction on the upper and lower boundaries, respectively (blue and red curves in Fig. 4 (a)). These features are consistent with the effective medium model. We also note that in the effective medium model the upper edge of this nontrivial band gap corresponds to $\mu_{\text{eff}} = 0$, while the lower edge corresponds to $\mu_{\text{eff}} \to \infty$. (See Supplemental Material Sec. I) This feature is consistent with Ref. [45]. However, the interaction in Ref. [45] is short-ranged (only limited to particles within a finite distance), our construction does not make assumptions about the ranges of interactions between particles, and the electromagnetic interaction between particles in the array systems similar to ours is typically not short-ranged.

The existence of one-way edge state in a random system can also be visualized by simulating a finite system as shown in Fig. 4(b). The cylinders are the same as Fig. 4(a) and the minimal distances between cylinders are also kept to be larger than $0.8a$. This lower-side of the finite system is truncated by a perfectly matched layer [46] to absorb the wave. The remaining boundaries of the system consist of PEC. The wave propagates anticlockwise without being backscattered when the frequency of the source is inside the nontrivial bandgap. While here we show an example of a random lattice, similar effects of band gap and one-way edge states are also observed in other isotropic systems such as quasicrystals with the same cylinder and cylinder density. (See Supplemental Material Sec. IV)

In conclusion, we theoretically propose and numerically demonstrate a route towards creating a metamaterial that behaves as a photonic Chern insulator, through homogenization of an array of particles. Using this route, we show that non-trivial topology can arise in a wide variety of lattices, including periodic, quasi-periodic and random lattices. While for concreteness we have considered cylindrical particle possessing gyromagnetic effects, one can achieve similar results with gyroelectric effects more commonly used in the optical frequency range. Also the particles can be of other shapes provided that their response can be well described by the electric and magnetic polarizabilities. Our method can easily be extended to create meta-materials that achieve electromagnetic analogues of quantum spin Hall systems, and be generalized for other classical wave systems. One can also imagine dynamic control of topological properties through the control of the densities of the particles.

This work is supported by the U. S. Air Force of Scientific Research (Grant No. FA9550-12-1-0471), and the U. S. National Science Foundation (Grant No. CBET-1641069).

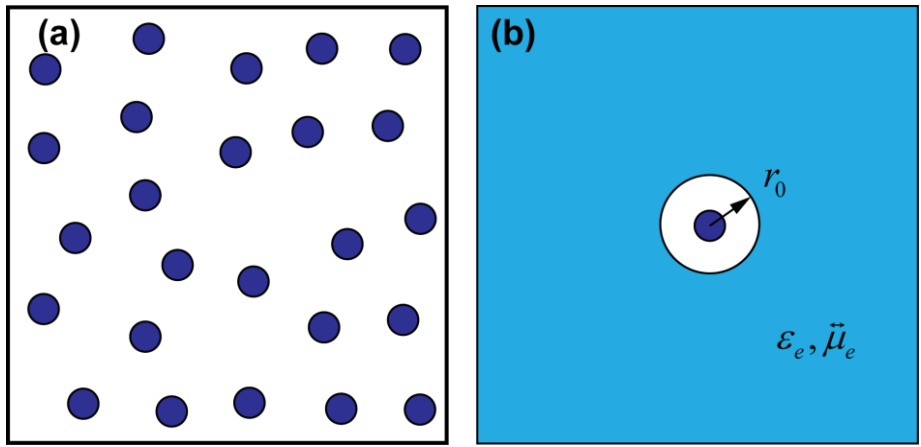

Fig. 1(color online). (a) A system consisting of a random array of cylindrical particles (solid blue disks). Each particle is described by its electric and magnetic dipole responses to external electromagnetic fields. (b) A sketch showing the geometry used in deriving the effective parameters, $\varepsilon_e$ and $\ddot{\mu}_e$, which provide an effective medium model for the array in (a). The light blue regions is filled uniformly with material having such effective parameters.

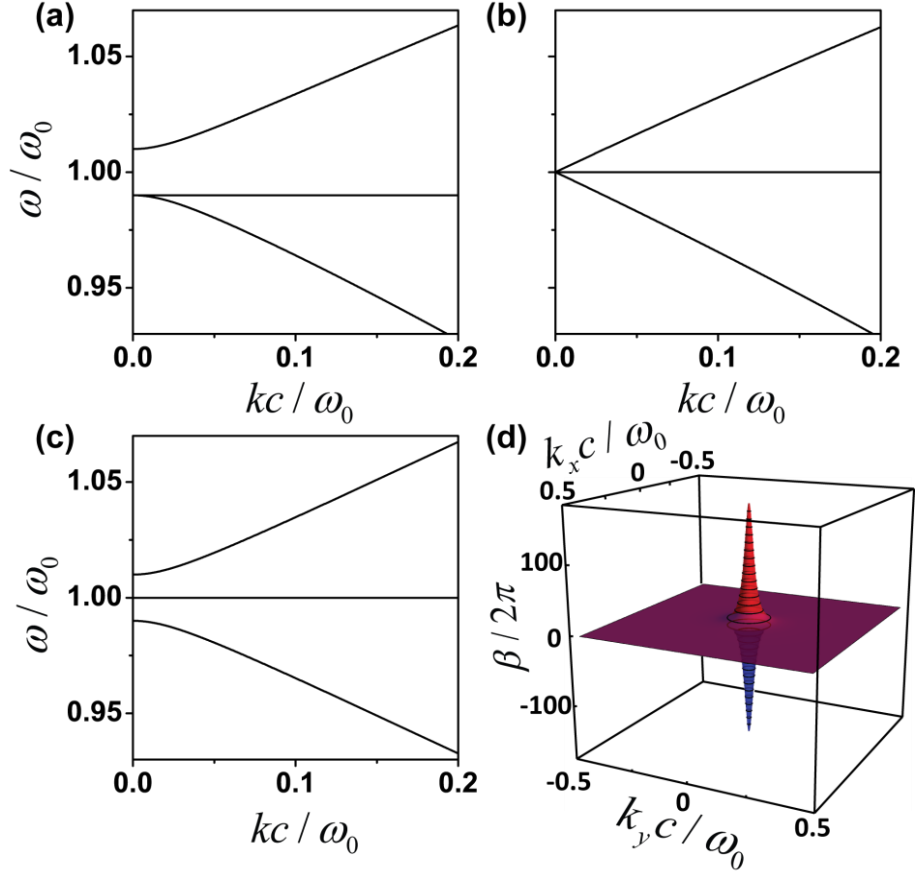

Fig. 2 (color online). The dispersion relation for the system as described by Eqs. (1)-(3) with $\gamma_\varepsilon = \gamma_\mu = 3$. $\omega_0$ is a reference frequency, $c$ is the speed of light in vacuum and $k$ is the in-plane wave vector. (a) $\omega_E = 1.01\omega_0$, $\omega_H = 0.99\omega_0$, $\mu_k = 0$. (b) $\omega_E = \omega_H = \omega_0$ and $\mu_k = 0$. (c) $\omega_E = \omega_H = \omega_0$ and $\mu_k = 0.03$. (d) Red and blue represent the Berry flux $\beta$ of the highest and lowest bands in (c), respectively. The Berry flux of the middle band in (c) is exactly zero.

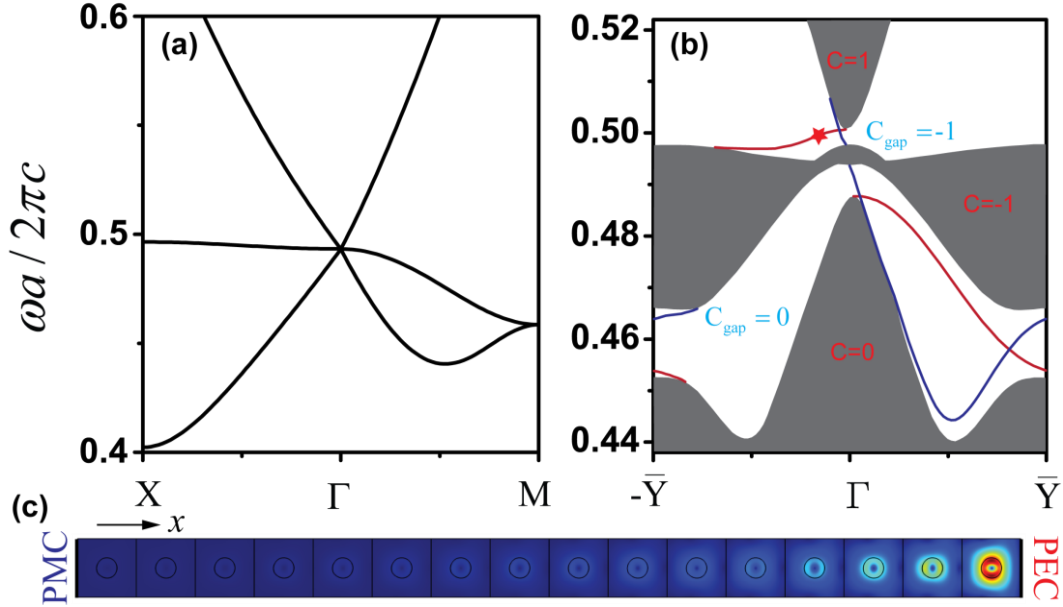

Fig. 3 (color online). (a) The band structure of cylinders in a square lattice. (b) Projected band structure (gray area) and corresponding surface states with the supercell in (c). This supercell is terminated by a PMC boundary on the left and PEC boundary on the right. Here red and blue curves represent the surface states at the PEC and PMC boundaries, respectively. The Chern numbers for the band and the gap Chern numbers in (b) are labeled in red and cyan, respectively. The electric field amplitude of the surface state marked by red star ($k_y a / \pi = -0.2$) in (b) is also shown in (c), where red and blue colors represent maximum and zero field amplitudes, respectively. The cylinder has a radius of $r_c = 0.1735a$, where $a$ represents the lattice constant of the square lattice. The relative permittivity of the cylinders is $\epsilon_c = 20$. $\kappa = 0$ in (a) and $\kappa = 0.08$ in (b) and (c).

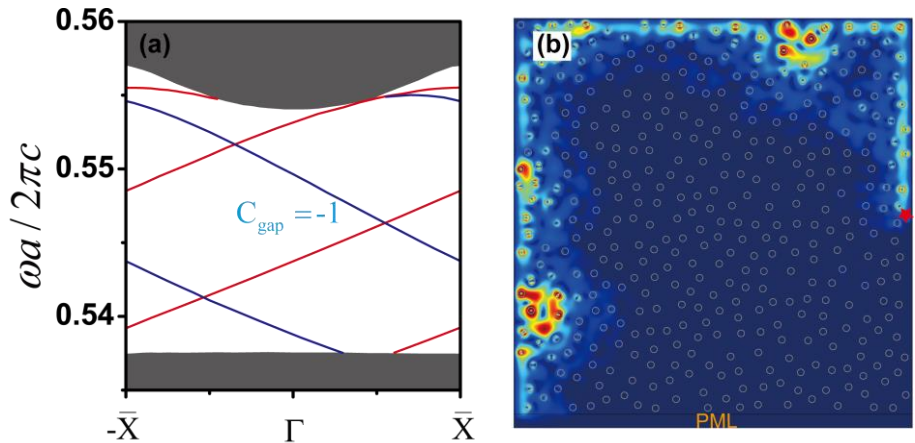

Fig. 4 (color online). (a) The projected bulk band structure (gray regions) and the band structure of surface states (red and blue curves) of a strip geometry consists of 5 of the supercells, each of which is shown in Fig. 1(a). The upper and lower boundaries of the strip are terminated by PECs and the strip is periodic along the horizontal direction. The blue and red curves represent the surface states on the upper boundary and the lower boundary, respectively. (b) One-way edge modes supported by a finite random lattice of cylinders. Here the red star marks the position of the source with an operating frequency at $\omega a/(2\pi c) = 0.547$. The rectangles on the lower end of the figure represents a perfectly matched layer (PML) region that absorbs the waves. All the other outer boundaries in (b) are PEC. The radius of the cylinder and the relative permittivity are $r_c = 0.1735a$ and 20, respectively, where $1/a^2$ is the density of the cylinders, and $\kappa = 0.4$.